\documentclass[useAMS,usenatbib]{mn2e}
\usepackage{mathrsfs} 
\usepackage{graphicx}
\usepackage{color}

\def\kms{km s$^{-1}$}  
\def\Msun{M_{\odot \hskip-5.2pt \bullet}} 
 \def\deg{^\circ}
\def\htwo{H$_2$} \def\vr{v_{\rm r}} 
\def\sin{{\rm sin}\ } \def\cos{{\rm cos}\ }
\def\N{N_{\rm HI}}  \def\muG{\mu{\rm G}} \def\nel{n_{\rm e}}
\def\Emag{E_{\rm mag}\ }
\def\apj{ApJ} \def\mnras{MNRAS}  
\def\RMunit{ rad m$^{-2}$ }
\def\xe{x_{\rm e}}\def\Ne{N_{\rm e}}\def\NH{N_{\rm HI}}  
\def\Hcmcm{{\rm H~ cm^{-2}}}  
\def\rperp{r_{\rm \perp}}  
\title[]{Three-Dimensional Aquila Rift: Magnetized HI Arch Anchored by Molecular Complex}  

\author[]{Yoshiaki Sofue$^{1}$\thanks{E-mail:sofue@ioa.s.u-tokyo.ac.jp} and  Hiroyuki Nakanishi$^{2,3}$ \\
$^{1}$ Insitute of Astronomy, The University of Tokyo, Mitaka, Tokyo 181-0015, Japan \\
$^{2}$ Graduate Schools of Science and Engineering, Kagoshima university, 1-21-35 Korimoto, Kagoshima 890-8544, Japan\\
$^{2}$ Institute of Space and Astronautical Science, JAXA,
3-1-1 Yoshinodai, Sagamihara, Kanagawa 252-5210, Japan
}

\begin{document} 
\date{}
   
\maketitle   
 
 \begin{abstract}
Three dimensional structure of the Aquila Rift of magnetized neutral gas is investigated by analyzing HI and CO line data. The  {  projected distance on the Galactic plane} of the HI arch of the Rift is  $\rperp \sim 250$ pc from the Sun. The HI arch emerges at $l\sim 30\deg$, reaches to altitudes as high as $\sim 500$ pc above the plane at $l\sim 350\deg$, and returns to the disk at $l\sim 270\deg$. The extent of arch at positive latitudes is $\sim 1$ kpc and radius is $\sim 100$ pc. The eastern root is associated with the giant molecular cloud complex, which is the main body of the optically defined Aquila Rift.  The HI and molecular masses of the Rift are estimated to be $M_{\rm HI}\sim 1.4\times 10^5\Msun$ and $M_{\rm H_2}\sim 3\times 10^5\Msun$. Gravitational energies to lift the gases to their heights are $E_{\rm grav: HI}\sim 1.4\times 10^{51}$ and $E_{\rm grav: H_2}\sim 0.3\times 10^{51}$ erg, respectively. Magnetic field is aligned along the HI arch of the Rift, and the strength is measured to be $B\sim 10\ \muG$ using Faraday rotation measures of extragalactic radio sources. The magnetic energy is estimated to be $E_{\rm mag}\sim 1.2\times 10^{51}$ erg. A possible mechanism of formation of the Aquila Rift is proposed in terms of interstellar magnetic inflation by a sinusoidal Parker instability of wavelength of $\sim 2.5$ kpc and amplitude $\sim 500$ pc.
\end{abstract}
 
\begin{keywords}
galaxies: individual (Milky Way) --- ISM: HI gas --- ISM: molecular gas --- ISM: magnetic field --- Parker instability
\end{keywords}
 
\section{Introduction}  

The Aquila Rift is a giant dark lane dividing the Milky Way by heavy starlight extinction at galactic longitude $l\sim 10-30\deg$ near the galactic plane (Weaver 1949; Dame et al. 2001; Dobashi et al. 2005), where it makes a giant triangular shade. It then extends to positive latitudes over the Galactic Center, and returns to the galactic plane at $l\sim 270\deg$ (Dame et al. 2001). It is composed of neutral gas and dust drawing a giant arch on the sky. 

Similarly to the dark lane, the HI arch is inflating from the galactic plane at $l\sim 30\deg$, extends over the Galactic Center and returns to the galactic plane at $l\sim 270\deg$, spanning over $\sim 120\deg$ on the sky. It also extends toward negative latitudes from $l\sim 30\deg$, running through $(60\deg,-40\deg)$ to $(90\deg,-50\deg)$.

 The Rift is composed of broad curved ridge of HI filaments (Kalberla et al. 2003). It is associated with a complex of dark clouds (Dobashi et al. 2003), molecular cloud complex (Dame et al. 2001) and dust emitting far infrared (FIR) emissions (Planck Collaboration et al. 2015a,b). Intensity cross section of the  HI arch and stellar polarization intensities indicates ridge-center peaked profile indicative of a filament or a bunch of filaments (Kalberla et al. 2005), but it does not show any signature of a shell. In this regard, the Aquila Rift is a filament or a loop, but not a shell, though it may be thought to be one of Heiles (1998) HI shells.

The Aquila Rift is associated with magnetic fields aligned along the ridge as inferred from star light polarization (Mathewson and Ford 1970; Santos et al. 2011) and FIR polarized dust emission (Planck collaboration 2015a,b).  
The eastern end of the HI arch near the galactic plane is rooted by another molecular cloud complex extending to negative galactic latitudes (Kawamura et al. 1999).

 The root region at $l\sim 20-30\deg$ apparently coincides with the root of the North Polar Spur (NPS). However, the NPS and Aquila Rift are shown to be unrelated objects separated on the line of sight from their different distances and perpendicular magnetic fields (Sofue 2015: See discussion section and the literature therein).

 The Aquila Rift has been modeled as a loop produced by interaction of the local hot bubble and Loop I (Egger and Aschenbach 1995; Heiles 1998; Reis and Corradi 2008; Santos et al. 2011; Vidal et al. 2015), and the loop is thought to be one of the super HI shells of Heiles (1984). This loop model has recently encountered a difficulty about its relation to the North Polar Spur (NPS; Loop I), because the NPS distance has been measured to be farther than several kpc by soft-Xray absorption (Sofue 2015; Sofue et al. 2016; Lallemant et al. 2016)) and Faraday rotation and depolarization analyses (Sun et al. 2014; Sofue 2015). Thus, the loop model seems to have lost its phenomenological basis.  Another difficulty is the origin of the loop, which has nothing special in loop center. 
 
  Therefore, it would be worth visiting the Aquila Rift from a quite different stand point of view based on more realistic astrophysical consideration about the formation mechanism such as the Parker instability in the magnetized interstellar medium, which is well established and widely accepted in the astrophysics of the galactic disk. Thereby phenomenologically,  besides the shells and loops, we may more carefully inspect into the HI, CO and magnetic field maps of the local ISM with particular insight into coherent filamentary structures.

In this paper we analyze the kinematics and morphology of interstellar HI and \htwo ~ gases in the Aquila Rift using the Leiden-Argentine-Bonn all-sky HI line survey (Kalberla et al. 2005 ) and Colombia galactic plane CO line survey (Dame et al. 2001). We present the three-dimensional structure of neutral gas distribution based on the result of kinematical analysis. We further discuss the magnetic field structure, and consider the origin in terms of the Parker instability. 

\section{Kinematical Distance of the Aquila Rift}  

\begin{figure}\begin{center}
\includegraphics[width=8cm]{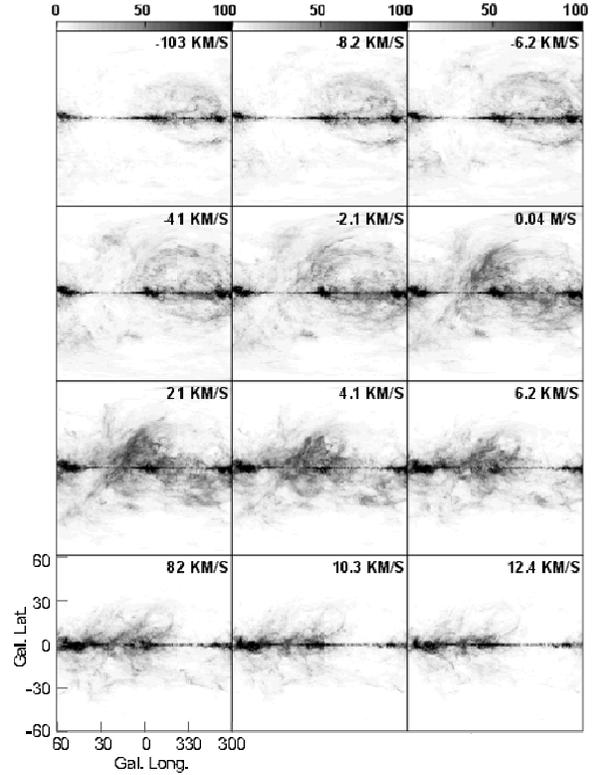} 
\end{center}
\caption{HI channel maps from LSR velocity $\vr= -10.3$ to 12.4 \kms from Kalberla et al. (2005). Intensity scale is HI brightness temperature in K indicated by the bar.}
\label{HIchannel} \end{figure}

\begin{figure}\begin{center} 
\includegraphics[width=8cm]{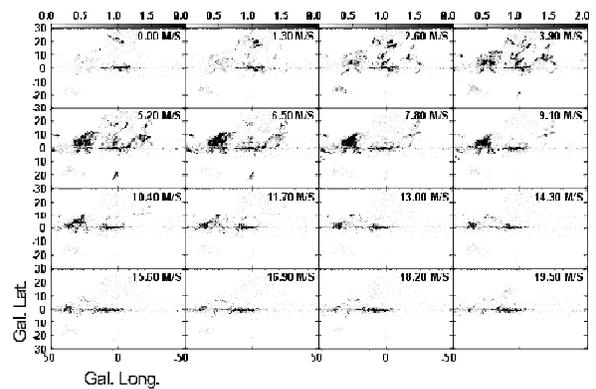}  
\end{center}
\caption{CO-line velocity channels map from LSR velocity, $\vr= 0$ to 19.5 \kms from Dame et al. (2001). Note the data at higher latitudes than $b>7\deg$ are not complete. Intensity scale is CO brightness temperature in K indicated by the bar.}
\label{COchannel}\end{figure}

 Figures \ref{HIchannel} and \ref{COchannel} show LSR velocity channel maps of Aquila Rift region. The figures show that the Aquila Rift is composed of HI and CO gases with radial velocities from $\sim -2$ to 10 \kms.
 The centroid velocity of HI ridge is around $\vr \sim 3$ \kms, while that for CO emission is at $\vr \sim 7$ \kms, systematically displaced from each other.  This indicates that the HI ridge is closer than the CO complex.
  
\begin{figure}\begin{center}
\includegraphics[width=8cm]{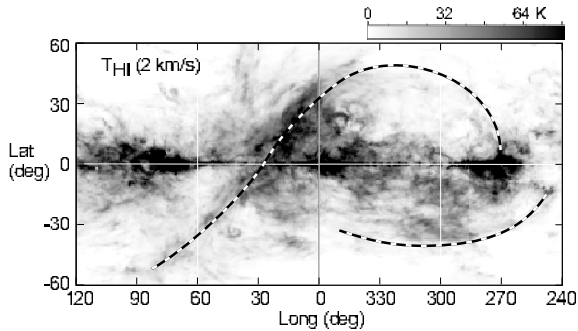} 
\end{center}
\caption{Enlagement of HI channel map at $\vr=2$ \kms. The sinusoidal dashed line traces the HI arch of the Aquila Rift, which is analyzed in this paper. The shorter dashed arc at right-bottom traces a possible counter arch. }
\label{enlargeHI} \end{figure}

In figure \ref{enlargeHI} we show an enlarged channel map at $\vr=2$ \kms, and an integrated intensity map from $\vr=-2$ to 6 \kms. Among the numerous filaments, the most prominent structure is the Aquila Rift traced by the sinusoidal dashed line. The main ridge runs from $(l,b)\sim(90\deg,-60\deg)$ through $(30\deg,0\deg)$ to $(0\deg,30\deg)$, reaching $\sim (270\deg,0\deg)$. The open structure of the Aquila Rift is evident in these maps.

\begin{figure}\begin{center}  
\includegraphics[height=7.5cm]{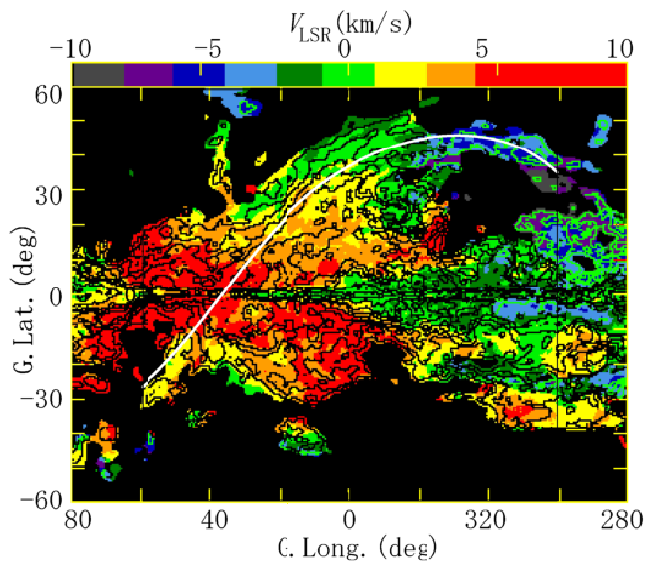}  
\end{center} 
\caption{HI velocity field (moment 1) in color, showing the general galactic rotation of the local HI gas by changing color from red to blue about $l=0\deg$. Overlaid are contours of integrated intensity from $\vr=-10$ to $+10$ \kms (moment 0) at every 50 K \kms.  
The inserted curve shows a sinusoidal fit of the Aquila Rift by parameters listed in table \ref{tabpara}. (color only in online journal)}
\label{vfield}\end{figure}

Another prominent arch is found in the south as a horizontal arc running through $\sim(300\deg,-33\deg)$. This 'counter' arch in the south is a separated structure from the northern arch of the Aquila Rift. In fact its radial velocity is significantly different from the Aquila Rift as shown by a velocity field in figure \ref{vfield}. 

Thus, the HI Aquila Rift is traced here as an open structure, drawing a half sinusoid on the sky, as shown by the dashed line in figure \ref{enlargedHI}. In the following analysis we focus on this northern Aquila Rift.

In figure \ref{vfield} we show a color-coded velocity field (moment 1) overlaid by a contour map of the integrated intensity (moment 0) from $v=-10$ to $+10$ \kms with the cut-off brightness temperature at $T_{\rm B: cut}=10$ K. The velocity field is smooth, and systematically changes the sign from negative to positive around $l\sim 0\deg$, indicating the general galactic rotation.

\begin{figure}\begin{center} 
\includegraphics[width=8.5cm]{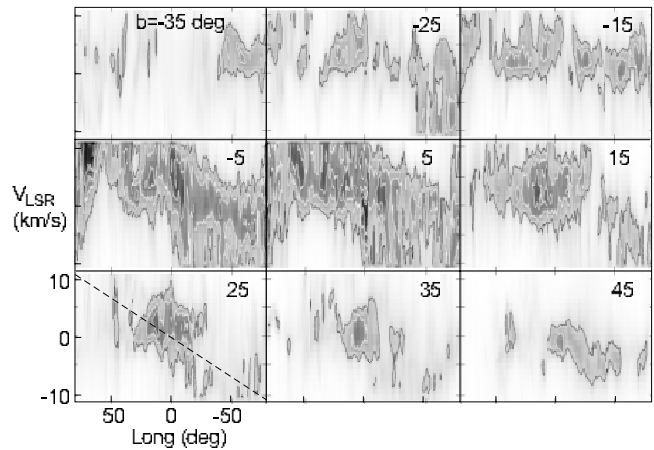}   
\end{center} 
 \caption{HI longitude-velocity (LV) diagrams at $b=-35\deg$ (top-left) to $+45 \deg$ (bottom-right) by every $10\deg$. 
The coherently tilted LV ridges at all latitudes indicate that the HI gas in the Aquila Rift region obeys the general galactic rotation. 
Contours are at 20, 40, 60, 80 K in brightness temperature. The dashed line represents the LV gradient used for distance determination by the $dv/dl$ method.}
\label{lvdiagram}\end{figure}

In figure \ref{lvdiagram} we show longitude-velocity (LV) diagrams at different latitudes. The mean velocity of the HI gas increases from negative to positive smoothly as the longitude increases, showing again that the gas is rotating with the galactic disk.   

The  {  distance projected on the galactic plane, $r$,} of a local object is given by    
\begin{equation}
r={\vr \over{A ~ {\rm sin}~2l }~ {\rm cos}~b }.
\label{vr_r}
\end{equation}
Here, $A$ is the Oort's $A$ constant, and we adopt the IAU recommended value $A=14.4$ \kms kpc$^{-1}$. 

This formulation results in large errors in distances near $l\sim 0\deg$. So, we did not use data at $|l|< 15\deg$ in order to avoid this. We also rejected HI data at $|b|<7\deg$, and CO data at $|b|<5\deg$ in order to avoid confusion with the disk component. We also avoided data with forbidden velocities for which the kinematical distance method cannot be applied. 

We here apply an alternative and more relible method to measure the distance to objects in the Galactic Center direciton, which was developed for distance determination of spiral arms in the GC direction (Sofue 2006). By this $dv/dl$ method, the projected distance $\rperp$ is given by
\begin{equation}
\rperp={R_0 \over V_0} {d\vr \over dl}=1.93 {d\vr \over dl\deg}~{\rm kpc}.
\label{dvdl}
\end{equation}
Here, $R_0=8$ kpc and $V_0=238$ \kms are the solar constants, and $\vr$ and $l$ are measured in \kms and degrees, respectively. The method is not sensitive to arm's radial, non-circular, or parallel motions. 

From figure \ref{lvdiagram}, the velocity gradient around $l\sim 0\deg$ is measured to be $d\vr/dl\simeq 0.13\ {\rm km\ s^{-1}}/{\rm deg}$ as indicated by the dashed line in the figure. Inserting this gradient, we obtain the  {  projected distance of the Aquila Rift as  $\rperp\simeq 250$ pc.} We adopt this value in the following analyses. We also confirm that the thus determined distance is consistent with the kinematical distance of the HI ridge at $|l|>\sim 15\deg$.  {  The actual distance to the Aquila Rift at its nearest point about $(l,b)\sim (20\deg,20\deg)$ is estimated to be  $d \sim 300$ pc.}

\section{3D Structure}

\begin{figure}  \begin{center} 
\includegraphics[width=8cm]{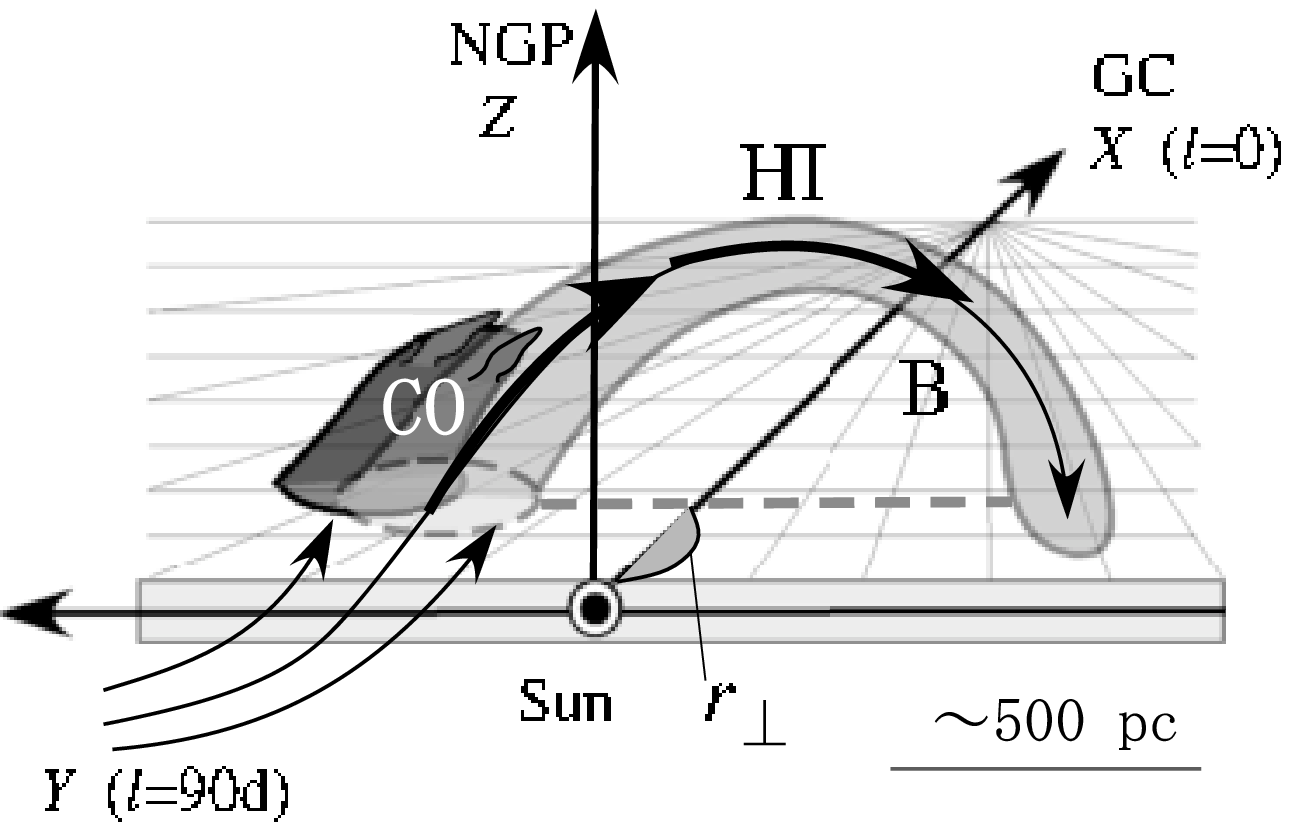}   
\end{center}
\caption{Coordinates used in the analysis, and illustration of 3D structure of the Aquila Rift. }
\label{illust}\end{figure}

\begin{figure*} 
\begin{center} 
\includegraphics[width=12cm]{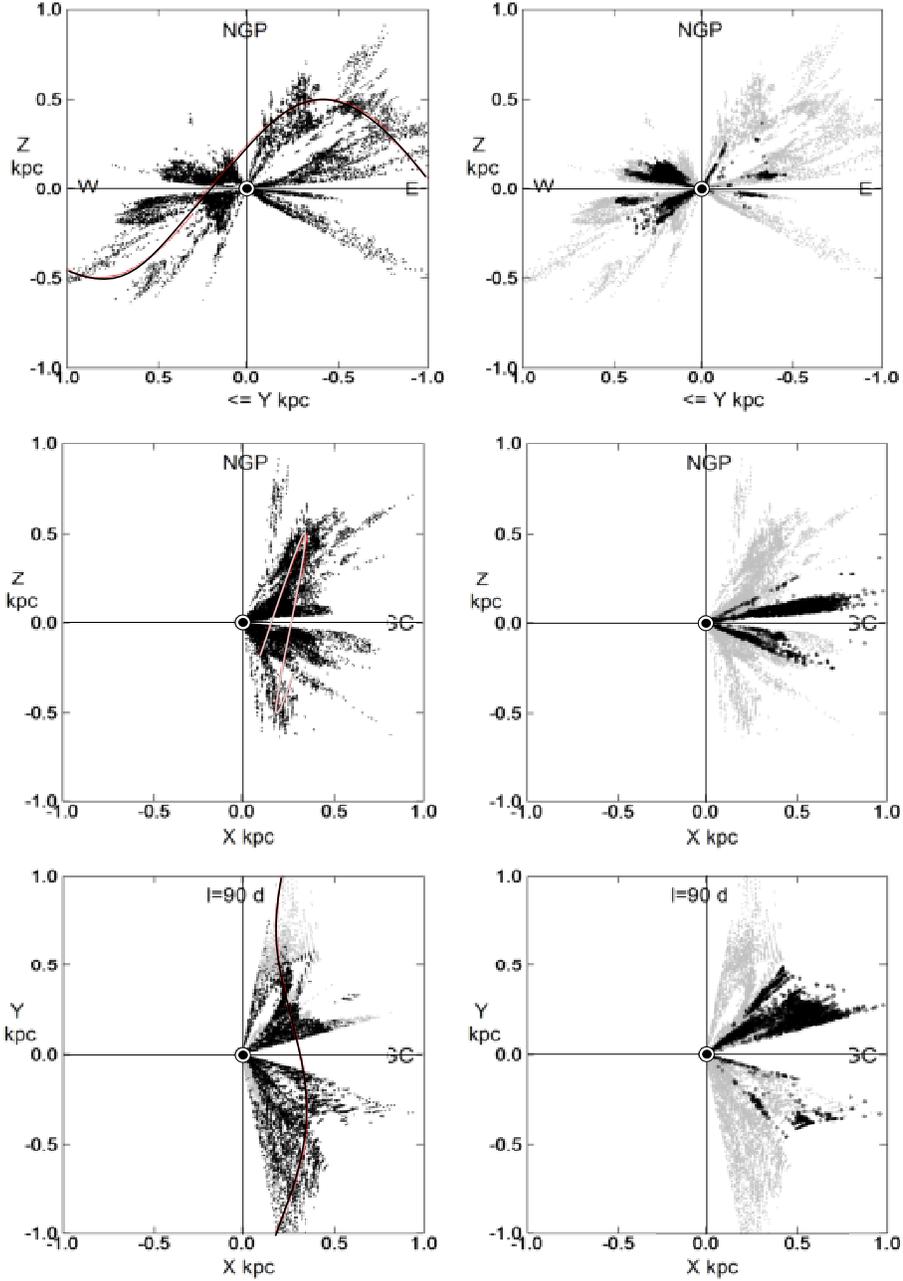}    
\end{center}
\caption{{\it Left panels}:  Projected positions of measured HI cells in the Aquila Rift with brightness temperature higher than 10 K. Data with negative-$Z$ in the $XY$ projection are shown by gray dots. The sinusoidal curve approximately fits the ridge of the HI arch by equations \ref{sinz} and \ref{sinx}. To avoid large kinematical distance errors, regions at $l>90\deg$, $15\deg>l>-15\deg$, and $l<280\deg$ were not used, whose bountaries appear as the straight cuts in the plots.
{\it Right panels}: Same, but for H$_2$ by CO line data with temperature higher  than 0.1 K. HI positions are superposed by gray dots.}
\label{xyz}
\end{figure*}

We construct a 3D map of the Aquila Rift transforming the $(l,b,\vr)$ cube data to those in an $(X,Y,Z)$ Cartesian coordinates as defined in figure \ref{illust}, where  $X=r \ \cos b \ \cos l$ and $Y=r \ \cos b \ \sin l$ and  $Z=r \ {\rm sin}\ b$. In figure \ref{xyz} we plot the HI and CO cell positions in the ($X,Y,Z$) coordinates, at which the HI brightness temperature was observed to be higher than the threshold values, which were taken to be 10 K and 0.1 K for HI and CO, respectively.

We created moment maps from the data cubes 
 {  by cutting data with temperatures below 10 and 0.1 K for HI and CO, respectively, representing integrated intensity $I$ from $\vr=-10$ to 10 \kms, mean velocity $v$, and velocity dispersion $\sigma_{v}$. If we assume that the Aquila Rift has a single velocity component, which is the case from the observed line profiles, the line width may be approximated by the dispersion.
}

 We thus obtained a pseudo-brightness temperature at $v$ as
$T_i(v)={I / \delta v}.$ 
The volume gas density is then calculated by 
\begin{equation}
n_i(r)={dN_i \over dr}=A C_i T_i(v)\ {\rm sin~}2l~ {\rm cos}~b,
\label{density}
\end{equation}
where $C_i$ are the conversion factors for HI and CO line intensities,
$
C_{\rm HI}=1.82\times 10^{18} {\rm H~cm}^{-2}({\rm K~km^{-1}})^{-1}
$
 and 
$
 C_{\rm H2}=2.0\times 10^{20} {\rm H_2~ cm}^{-2}({\rm K~km^{-1}})^{-1}.$
 
\begin{figure*} 
\begin{center}
\includegraphics[width=12cm]{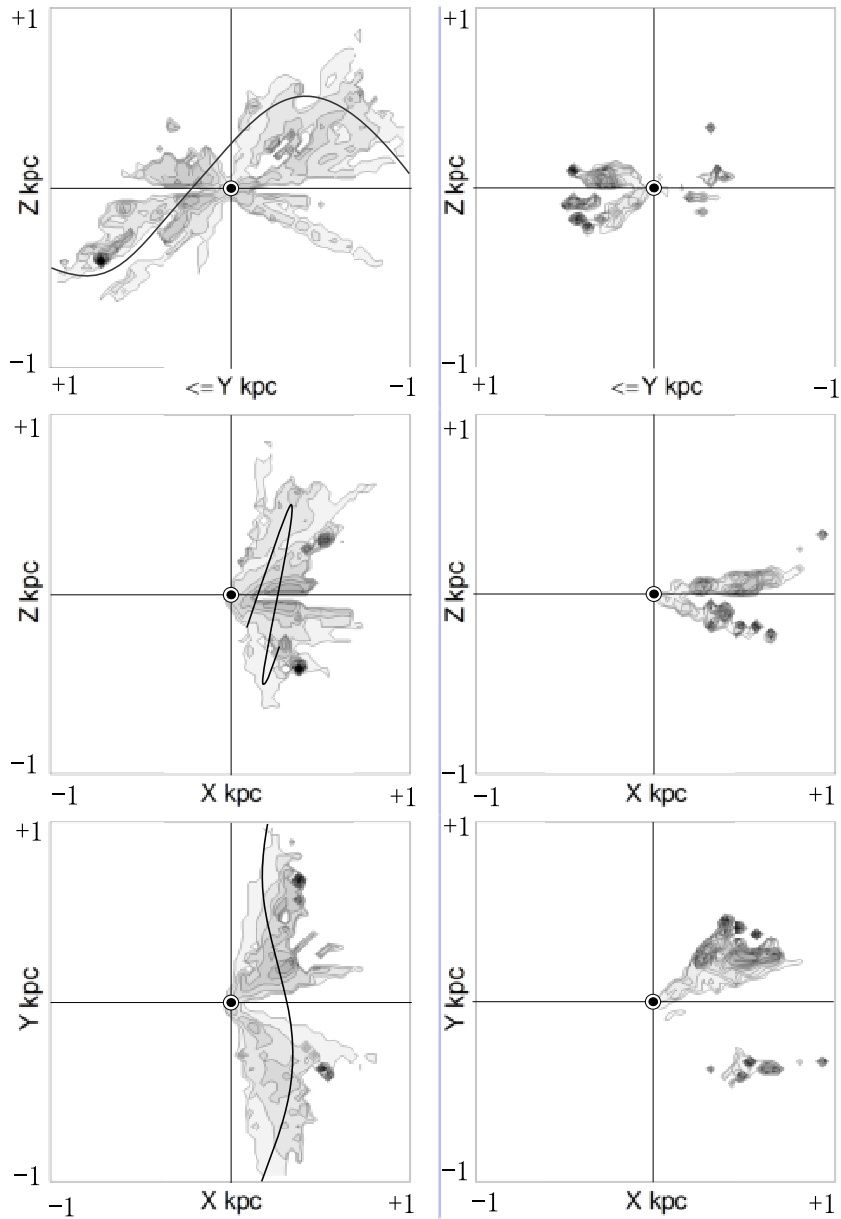}     
\end{center}
\caption{{\it Left}: HI volume density averaged in the directions perpendicular to the Cartesian planes. Contours are at 2.5, 5, 7.5, ... H cm$^{-3}$. {\it Right}: Same, but for H$_2$. Contours are at 5, 10, 15, .... H$_2$ cm$^{-3}$. }
\label{den_xyz}
\end{figure*}

 {  Relating $(r,l,b)$ to $(X,Y,Z)$, we finally obtain $n_i(X,Y,Z)$. This density is usually much higher than the mean density calculated from moment 0 map.}
Figure \ref{den_xyz} shows the obtained distributions of HI and H$_2$ gas densities projected on the Cartesian planes.   

The 3D density distribution shows that the HI Aquila Rift is extended over a length of $L\sim 2$ kpc in the $Y$ direction. The height of the top is about $Z\sim 500$ pc, and the width is approximately $\sim \pm 100$ pc. 
 {From the 3D plots combined with the arched appearence shown in section 2, the HI arch may be roughly represented by a sinusoidal curve, as drawn in the 3D figures. 
Here, the the approximate $X$ displacement is $\sim 0.25$ kpc, $Y$ displacement of the node $\sim 0.2$ kpc, and amplitude $\sim 500$ pc. The entire arch is tilted from the $Y$ axis by $\sim 5\deg$ toward $l\sim 85\deg$ and from the vertical plane by $\sim 15\deg$ toward GC.  }

In figures \ref{xyz} and \ref{den_xyz} are also shown the derived 3D CO maps.
The kinematical distance to the CO Aquila Rift has been estimated to be $640\pm 170$ pc in our previous work (Sofue 2015). The height of the cloud center is  $z\sim 60$ pc, and the linear extent about 100 pc elongated in the line of sight direction. The here obtaind 3D maps are consistent with these estimation, as the data and method are the same.  

 {  The 3D structure of the entire Aquila Rift was thus obtained for the first time, displaying the HI and H$_2$ gas densities in the Cartesian coordinate system. Without such 3D information with, we can neither calculate magntic field strength, nor can compare with the numerical simulations of the Parker instability, as discussed later.}
  
\section{Magnetic Arch} 

\subsection{Field orientation}

\begin{figure} 
\begin{center} 
\includegraphics[width=7cm]{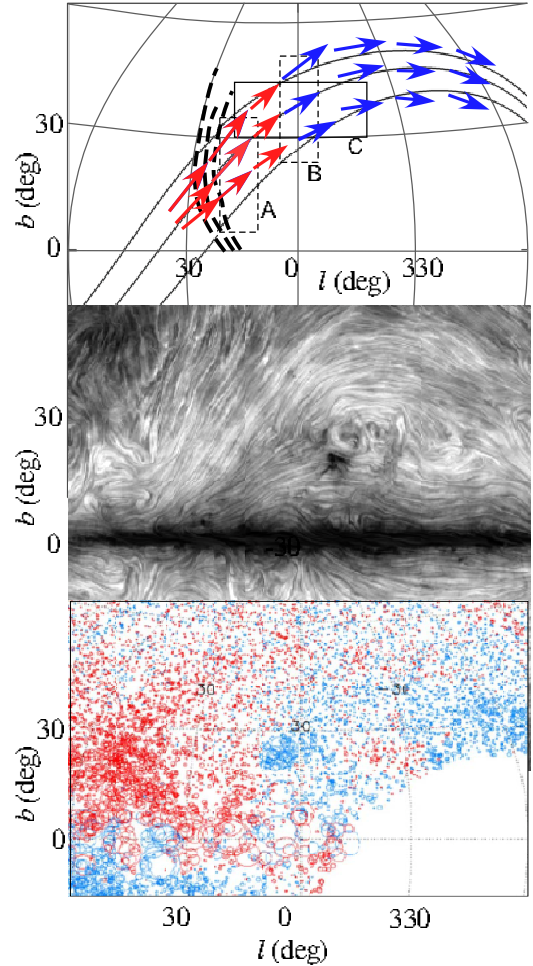}    
\end{center}
\caption{[Top] Arrows trace flow lines of magnetic field from dust emission (middle panel), with the directions inferred from RM signs (bottom). Areas A and B are regions for $dRM/dN$ analysis, and C for $dRM/d\xi$ analysis.  The sinusoidal curves show a model given by equations \ref{sinz} and \ref{sinx}. The dashed curves show magnetic direction of the North Polar Spur from polarized synchrotron emission (WMAP: Bennett et al. 2013).
 { [Middle]} Magnetic flow lines by polarized dust emission for $\pm 60\deg$ region around the GC (Planck collaboration 2015a,b). 
 { [Bottom]} Faraday RM from Taylor et al. (2009) for the same region, red denoting positive (field toward the observer) and diameter proportional to $|RM|$, and blue for negative (away from observer).  {(See the original papers for detail.)}}
\label{mag}
\end{figure}

Starlight polarizations have shown that the magnetic fields in the Aquila Rift are parallel to its ridge (Mathewson and Ford 1970). Figure \ref{mag} shows flow lines of the magnetic fields derived by polarization measurement of FIR emission of interstellar dust associated with the Aquila Rift, indicating that the field lines run along the main arc ridge (Planck collaboration 2015b). 

The line-of-sight direction of the magnetic field can be obtained from the distribution of Faraday rotation measure (RM). RM values observed for extragalactic radio sources (Taylor et al. 2009) are indicated by circles in the figure, where the diameter of a circle is proportional to $|RM|$. RM are positive at the eastern half of the arch at $l>\sim 10\deg$ indicating that the line-of-sight field direction is toward the observer, while they are negative in the western half at $<10\deg$ showing field running away from the observer.

Considering the 3D structure of the HI ridge of the Aquila Rift, we may draw the apparent direction of the magnetic field projected on the sky as indicated by the arrows in figure \ref{illust}. In figure \ref{mag} we draw three lines corresponding to $Z_1=-100,\ 0$ and 100 pc, mimicking sinusoidal magnetic lines of force about the galactic plane as simulated by a non-linear simulation of the growth of Parker instability (Matsumoto et al. 2009).  

\subsection{Field strength by $d RM/d\N$ method} 

\begin{figure} 
\begin{center} 
\includegraphics[width=8cm]{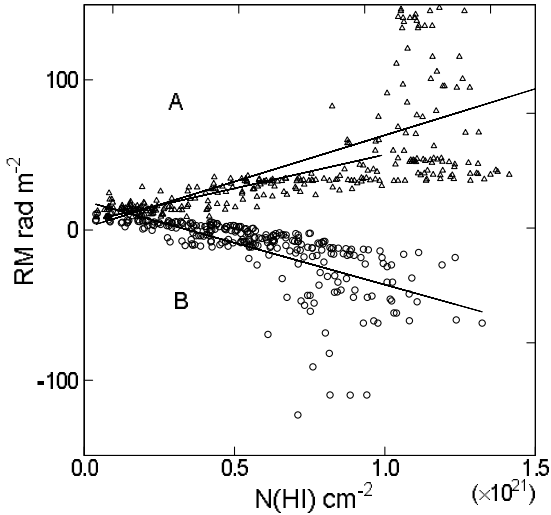}   
\end{center}
 {\caption{Rotation measure plotted against HI column density in two $10\deg\times 30\deg$ regions A (triangles) and B (circles) centered on $(l,b)=(20\deg,20\deg)$ and $(0\deg,40\deg)$, respectively, in figure \ref{mag}. The straight lines show the least-squares fitting results for A and B regions. The shorter line is a fit at $\N \le 10^{21}\ \Hcmcm$. } }
\label{HIRM}
\end{figure}

 {Figure \ref{HIRM} shows RM values plotted against HI column density for integrated intensity between $-10$ and 10 \kms in two regions along the Aquila Rift shown in figure \ref{mag}. 
The RM values were taken from the archival fits data by Taylor et al (2009) showing median RM values in $8\deg$ ($\sim$ FWHM) diameter circles, but regridded to a map of the same grids as the HI map ( FWHM$\sim 0\deg.6$) from Kalberla et al. (2005). } 
The plotted dots represent RM and $\NH$ values of individual mesh points.

The plot for region A centered at $l=20\deg$ and $b=20\deg$ show positive correlation, whereas that for region B in the western side centered on $l=0\deg$ and $b=40\deg$ shows negative correlation.  

If the magnetic field is ordered, the rotation measure is simply proportional to the column density of thermal electrons $\Ne$ as $RM\propto B \Ne$ and $\propto \Ne$ for constant $B$. On the other hand, if the magnetic fields are random and frozen in the gas, we have $B\propto \nel^{2/3}$ so that $RM\propto \Ne^{5/3} L^{-2/3}\nu^{-1/2}$, where $\nu$ is the number of gas eddies in the depth $L$, or we have $RM\propto \Ne^{5/3}$ for fixed $L$ and $\nu$. Although it is difficult to discriminate which is the case from the plots, we may here consider that the field is ordered along the arch from the regular flow lines of the dust polarization, and assume that $RM \propto \Ne$. 

The rotation measure is expressed by 
\begin{equation}
RM = 0.81 \int \nel B \cos \theta dL,
\label{RMdef}
\end{equation}
where $B$ is in $\muG$, $\theta$ is the viewing angle of the field lines, $\nel$ is the electron density in cm$^{-3}$, and $L$ is the line of sight distance in pc. This equation can be rewritten as
\begin{equation}
RM \simeq \Gamma x_{\rm e}  \N B\ \cos \theta,
\label{RM}
\end{equation}
where $\Gamma =2.62 \times 10^{-19}$ \RMunit cm$^2 \ \muG^{-1}$, $\N$ is the HI column density in cm$^{-2}$, and  $x_{\rm e}$ is the free-electron fraction. 

We here approximate the RM dependence on $\N$ by a linear relation 
\begin{equation}
RM =Q \N+{\rm const.}
\end{equation}
 with 
 \begin{equation}
 Q={dRM \over d\N}
 \end{equation}
 being the gradient of the plot. The magnetic field strength is then obtained by measuring the gradient $Q$ as
\begin{equation} 
B \simeq 3.82\times 10^{18} {Q \over  x_{\rm e} \cos \theta} \ \muG,
\label{eqB}
\end{equation} 
with $Q$ measured in \RMunit cm$^{2}$.

The free-electron fraction $x_{\rm e}$ is not measured directly in this particular region for the Aquila Rift. The fraction is generally determined by simultaneous measurements of dispersion measure and HI absorption toward pulsars, and is on the order of $x_{\rm e} \sim 0.08$ (Dalgarno and McCray 1972). About the same values have been obtained by UV spectroscopy of stars for the warm neutral hydrogen in the local interstellar space (Jenkins 2013), and by comparison of RM with HI column densities (Foster et al. 2013). We here adopt $x_{\rm e}\sim 0.08$ for the Aquila HI arch.

The $Q$ value is determined by linearly fitting the $RM-N$plot. We thus obtain
  $Q\sim 62\pm0.1\times 10^{-21}$ rad m$^{-2}{\rm cm}^{2}$ 
in region A, and 
  $Q\sim -56\pm 0.2\times 10^{-21}$ rad m$^{-2}{\rm cm}^{2}$ 
in region B. Here, A and B are $10\deg\times 30\deg$ squared areas centered on $(l,b)=(20\deg,20\deg)$ and $(0\deg,40\deg)$. They were so chosen to represent typical regions with positive and negative RM crossing the arch. The fitting results are shown by straight lines in the figure.

The viewing angles of the field lines toward the centers of regions were determined as the angles on the sky from the neutral region of $RM\simeq 0$ at $l\simeq 10\deg$ and $b\simeq 30\deg$, and are $\theta\simeq 81\deg$ and   $\simeq 102\deg $ for regions A and B, respectively.

For these measured values of $Q$ and viewing angles $\theta$, we estimate the field strength along the Aquila Rift to be 
$B \sim 12.5\ \muG$ and 
$B\sim 12.3\ \muG$ 
for regions A and B, respectively. Thus, the field strength along the arch is estimated to be $B\sim 12.4\ \muG$.

We comment that the $RM-\N$ plot includes high RM values at $\N>10^{21}$ H cm$^{-2}$, but we did not exclude them. If we remove such high RM data and use regions with mediate density regions of $\N \le 10^{21}$ H cm$~{-2}$, the field strength is reduced to $B\sim 6\ \muG$ for region A, while not changed in region B.

\subsection{Field strength by dRM/d$\xi$ method} 

Let us consider that we observe RM in region C around the perpendicular point of a magnetic tube as showin in figure \ref{mag}.  
Introducing an angle $\xi=90\deg-\theta$, equation \ref{RM} can be rewritten as
\begin{equation}
B\simeq{dRM\over d\xi}{1\over\Gamma \xe \NH},
\label{dRMdxi}
\end{equation}
for small $\xi$ around the perpendicular point to the field lines.
Measuring $B$ in $\muG$, $\NH$ in H cm$^{-2}$, RM in \RMunit, and $\xi$ in degrees, this equation is rewritten as
\begin{equation}
B\simeq  2.19 \times 10^{20} {dRM\over d\xi \deg}{1\over \xe \NH},
\end{equation}
or if $\xe=0.08$, we obtain
\begin{equation}
B\simeq  2.74 \times 10^{21} {dRM\over d\xi \deg}{1\over \NH},
\end{equation}
This method, called the $dRM/d\xi$ method, may be applied to the perpendicular region of magnetic field in the Aquila Rift. 

 { Figure \ref{dRMdxi} shows RM plotted against longitude
in region C ($-20<l<20\deg, 30<b<40\deg$) by grey dots, where RM values were obtained by the same way as explained in the previous subseciton. The figure shows a clear gradient of RM against the longitude. We further calculated averages in every $4\deg$ interval of longitude, and plot them by black dots with standard deviations by the bars. The plot may be well fitted by the inserted dashed line, to which we apply the present method. }

The RM varies from negative to positive at gradient of $dRM/dl\deg \simeq  {   1.0}$ \RMunit${\rm deg}^{-1}$. 
Correcting for the cos $b$ effect and tilt angle $\sim  {   30}\deg$ of the field line from a constant latitude, we estimate the gradient to be $dRM/d\xi \deg \sim  {   1.0}$ \RMunit${\rm deg}^{-1}$. The HI column density averaged in region C is $\NH\sim  {4}\times 10^{20}$ H cm$^{-2}$. Thus, the field strength in region C is obtained to be  $B\sim {  7}\ \muG$.

\begin{figure} 
\begin{center} 
\includegraphics[width=8cm]{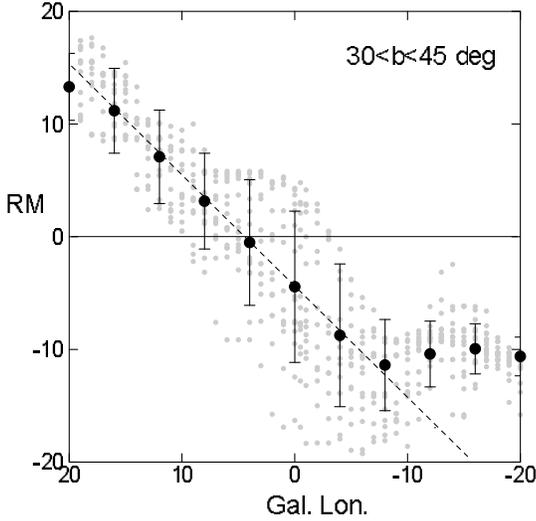}   
\end{center}
\caption{RM plotted against longitude near the perpendicular point  {in region C of figure 10. Grey dots are raw values from fits data by Taylor et al. (2009). Black circles are averages in longitude every 4$\deg$, and bars are standard deviation. The dashed line was used for $dR/d\Xi$ method.}
}
\label{dRMdxi}
\end{figure}

\section{Discussion}
 
\subsection{Assumption and limitation}

The kinematical distances derived from radial velocities using the Galactic rotation may include large uncertainty arising from non-circular motions such as turbulence. Although this is unavoidable, the symmetric velocity field about $l\simeq 0\deg$ and the LV diagrams indicate general galactic rotation of HI gas in the Aquila Rift.

 {  The projected distance to the Aquila Rift, $\rperp \sim 250$ pc, and the actual distance to the nearest point, $\sim 300$ pc, are not inconsistent with the scattered range of measured distances from the literature for various objects by various methods} 
(Mathewson and Ford 1979; Dzib et al. 2010; Puspitarini et al. 2014).
 The extent on the sky is wider than $\sim 120\deg$, and hence, the linear extent is greater than $\sim 2$ kpc. Since the extent is much larger than interstellar cloud sizes, we may consider that the Rift is a galactic structure rather than a turbulent interstellar cloud. 

 {  The assumption that the HI Aquila Rift has a single velocity component may make the model too simplified, only representing the backbone of the Rift.}
 The accuracy about the extent would be within a factor of $\sim 1.5$ inferred from the scatter of the currently measured distances in the literature, from $\sim 150$ pc  $\sim 400$ pc. This scatter may also affect the derived parameters such as the mass and energetics. 

Another effect that might affect the result is a possible vertical motion during the magnetic inflation. However, the velocity would be at maximum on the order of Alfv\'en velocity of a few \kms, and the line-of-sight velocity ($\propto \sin\ b$) is much less. 
 {  Morever, the $dv/dl$ method is little affected by such transverse motions.}

\subsection{HI and H$_{\rm 2}$ Masses} 

The HI brightness temperature along the ridge of the Aquila Rift is measured to be $T\sim 30\pm 15$ K, and the  full width of the HI line at half maximum (FWHM) is $\Delta v\sim 5$ \kms. Thus the averaged column density along the ridge is obtained to be $N=C_{\rm HI} T \Delta V \sim 3\times 10^{20}$ H cm$^{-2}$. An approximate mass of the HI gas in Aquila Rift may be calculated by $ m_{\rm H}NL_1 L_2\sim 2\times 10^5\Msun$, where $m_{\rm H}$ is the hydrogen atom mass.
A more accurate mass can be estimated by summing up the counts in the Moment 1 map (averaged intensity) for the Aquila Rift region, and we obtain the total mass of HI gas to be  $M_{\rm HI}\sim \sim 1.8 \times 10^5 \Msun$. 
 
From the CO-line intensities and linear extent, the molecular mass of the Aquila-Serpens molecular complex has been estimated to be $M\sim  3\times 10^5 \Msun$ for a conversion factor of $2\times 10^{20} {\rm H_2 cm^{-2} (K\ km\ s^{-1}})^{-1}$ (Sofue 2015). Taking a half velocity width $\sigma_v\sim 4$ \kms and radius $\sim 50$ pc, the Virial mass is estimated to be  $M_{\rm V} \sim 2 \times 10^5\Msun$. Thus, the complex is considered to be a gravitationally bound system, and may be one of the nearest giant molecular clouds (GMC). 
Table \ref{tabpara} lists the estimated quantities for the Aquila Rift.

\begin{table}
\caption{Parameters for the Aquila Rift. }
\begin{center}
\begin{tabular}{lll}  
\hline \hline
Parameter& HI arch & H$_2$ complex \\
\hline   
 {  Projected distance $\rperp$} & 250 pc &$400\sim 640$ pc\\ 
Length $L_1$  ($b>0\deg$)& $\sim 1$ kpc & $\sim 100$ pc \\
Total length $L$ & $\sim 2$ kpc \\
Width $L_2$ & $\sim 100$ pc & $\sim 100$ pc   \\
\hline
Mass & $1.8\times 10^5 \Msun $& $3 \times 10^5 \Msun$  \\
$E_{\rm grav}$ & $1.4\times 10^{51}$ erg& $0.3\times 10^{51}$ erg \\ 
$E_{\rm kin}$  & $< 0.1\times 10^{51}$ erg & $< 0.1\times 10^{51}$   \\ 
$B$ in A by $dRM/d\NH$ & $\sim 12.5\ \muG$ \\  
--- in  B ibid& $\sim 12.3\ \muG$ \\  
--- in  C by $dRM/d\xi$ & $\sim  {  7}\ \muG$ \\  
$E_{\rm mag}$ for $B\sim 10\ \muG $& $\sim 1.2 \times 10^{51}$ erg   \\ 
\hline
Sinusoidal fitting (Eq. \ref{sinz}, \ref{sinx}) \\
Wavelength $\lambda$ & $\sim 2.5$ kpc\\
$X$ displacement &$\sim 250$ pc\\
$Y$ displacement of root & $\sim 200$ pc\\ 
$Z$ height (amplitude)  & $\sim 500$ pc & 60 pc   \\ 
Tilt from $Z$ axis& $\sim 15\deg$ tow. GC \\
Tilt from $Y$ axis & $\sim 5\deg$ tow. $l=85\deg$ \\

\hline   
\end{tabular}
\end{center} 
\label{tabpara} 
\end{table}

\subsection{Energetics}

The gravitational energy to lift the HI gas to height $z$ is estimated by 
\begin{equation}
E_{\rm g}=\int M k_z dz\simeq M k_z z,
\end{equation}
where $k_z$ is the vertical acceleration. 
At the mean height of the HI arch of $z\sim Z_0/2\sim 250$ pc, the acceleration is $k_z\simeq 5\times 10^{-9}{\rm cm\ s^{-2}}$ (Cox 2000). Thus, the gravitational energy of the HI arch is estimated to be $E_{\rm g, HI} \sim 1.4\times 10^{51}$ erg.
That for the Aquila-Serpens molecular complex at mean height $\sim 60$ pc is estimated to be on the order of $E_{\rm g, H_2}\sim 0.3\times 10^{51}$ erg, where $k_z\simeq 3\times 10^{-9}{\rm cm\ s^{-2}}$ at $z\sim 60$ pc. 

The kinetic energy  of the HI arch and molecular complex is estimated to be $E_{\rm kin}<\sim 10^{50}$ erg for internal motions of several \kms at most, although their vertical motion is not measurable. 
 
The magnetic energy in the HI arch at positive latitude is estimated by assuming a magnetic tube of length $L_1\sim L/2$ and width $L_2$ with constant field strength as 
\begin{equation}
\Emag \simeq { B^2 \over 8\pi} L_1 L_2^2,
\end{equation}
Inserting the measured strength $B\sim 10\ \muG$ at $(l,b)\sim(20\deg,30\deg)$, $L_1\sim 1$ kpc and $L_2\sim 100 $ pc, we obtain the magnetic energy of $\Emag\sim 1.2\times 10^{51}$ erg, which is comparable to the gravitational energy.  
\subsection{Magnetized HI arch anchored by Molecular Complex formed by Parker Instability}

As to the origin of the Aquila Rift we may consider the Parker (1966) instability. As shown in table \ref{tabpara} the magnetic energy of the arch is comparable to the total gravitational and kinetic energy of the HI and H$_2$ gases. If we assume the pressure equilibrium between cosmic rays and magnetic field, the pressure is strong enough to raise the gas against the gravitational force, so that the condition for the growth of Parker instability is satisfied. 

There have been a number of linear and non-linear analyses of the Parker instability (Elmegreen 1982; Matsumoto et al. 1990; Hanawa et al. 1992) as well as 2D and 3D numerical simulations (Santill{\'a}n et al. 2000; Nozawa 2005; Mouschovias et al. 2009: Machida et al. 2009, 2013; Lee and Hong 2011; Rodrigues et al. 2016). Formation of arched magnetic structure associated with interstellar gas has been thus well understood theoretically. The simulations showed that the instability can grow either in symmetric or asymmetric waves with respect to the galactic plane. 

The latter may better explain the sinusoidal behavior of the Aquila Rift. In figure 13 we reproduce the magnetic field lines calculated by 2D MHD simulation by Matsumoto et al. (2009) and Mouschovias et al. (2009), superposed on intensity maps of local HI and CO gases. The linear scale is adjusted to fit the observed HI arch with $\lambda=2.5$ kpc, which scales the simulated result by Mouschovias et al. to thermal equilibrium height of disk to be 50 pc (in place of their original value of 35.5 pc).
 
\begin{figure}  
\begin{center}   
(a)\includegraphics[width=6.5cm]{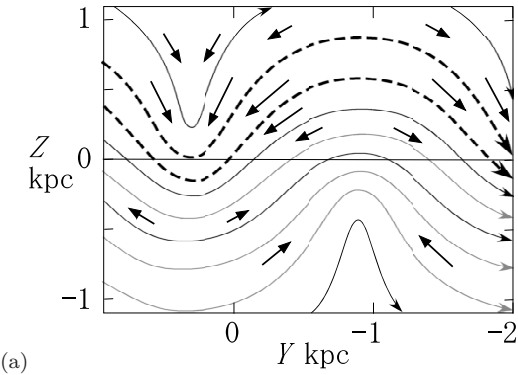}   
(b)\includegraphics[width=8cm]{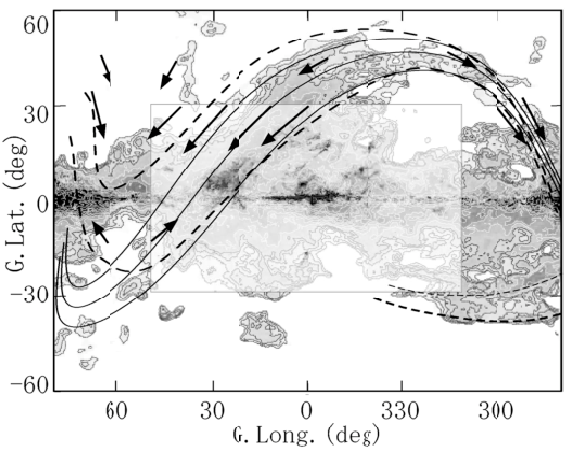}   
\end{center}
\caption{(a) Magnetic lines of force drawn after 2D simulation of Parker instability by Matsumoto et al. (1990) and Mouschovias et al. (2009)  {  scaled to wavelength $\lambda=2.5$ kpc}. Thick arrows indicate direction and speed of the gas flow. (b) Same as dashed curves in (a) projected on the sky overlaid on the intensity maps of local HI (grey + contours; figure 4; Kalberla et al. 2005) and CO (half-transparent; Dame et al. 2001) line emissions for tilt angles as in table \ref{tabpara}. Thin lines are the same but for sinusoidal field lines (Eq. \ref{sinz}, \ref{sinx}). 
The dashed curve at $(l,b)\sim(300\deg,-33\deg)$ is the southern coutner arch.
  {  The northern HI arch making the Aquila Rift and this southern arch may be two different flux ropes both affected by the same Parker instability.}
 }
\label{Parkerinsta}
\end{figure} 

\begin{figure} 
\begin{center} 
\includegraphics[width=8cm]{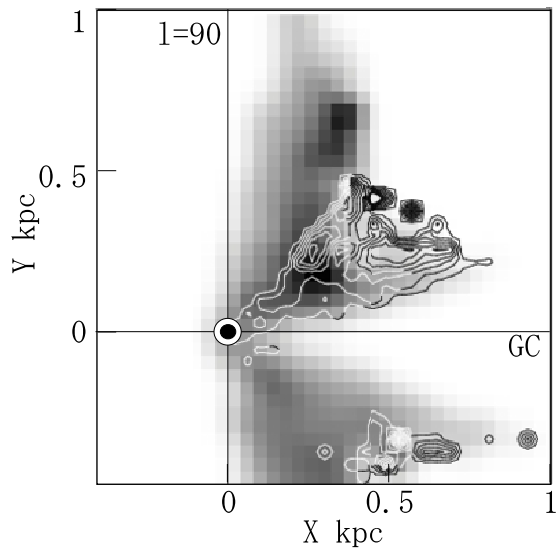}   
\end{center}
\caption{HI density grey-scale map compared with contour map of the molecular gas density projected on the $(X,Y)$ plane, showing that HI and molecular gases avoid each near the root of Aquila Rift.  See figure \ref{den_xyz} for intensity scales.}
\label{HIvsCO}
\end{figure}

The Aquila-Serpens molecular complex is located at the eastern root of the HI arch. This may be explained by accumulation of slipped-down HI gas from the top of the magnetic arch, where the HI gas is compressed to cause phase transition into molecular gas (Elmegreen 1982;  Mouschovias et al. 2009). It is interesting to notice that the HI and molecular gases avoid each other at the root of the Aquila Rift, as shown in figure \ref{HIvsCO}. The boundary of the two gases is sharp, as if the molecular complex is half-embedded by HI shell.

\subsection{Formation scenario of Aquila Rift}

We may summarize a possible scenario of formation of the Aquila Rift as follows (figures \ref{illust} and 13):\\
(1) The galactic disk about $\sim 250$ pc in the GC direction was penetrated by an ordered magnetic field parallel to the solar circle, and was compressed by a galactic shock wave to increase the density and magnetic field.\\
(2) A Parker instability took place along the shocked compressed high density arm with strong magnetic field, and the gas was inflated by an expanding magnetic tube to a height $\sim 500$ pc with a wavelength of $\sim 2.5$ kpc.\\
(3) It occurred in a time scale of $t\sim Z_0/v_{\rm A} \sim 5\times 10^7$ y, where $v_{\rm A}\sim 10$ \kms is the Alfv\'en velocity for $B\sim 10\ \muG$ and $n_{\rm H}\sim 5$ H cm$^{-3}$.\\
(3) The inflated HI gas slipped down along the field lines toward the Aquila region, where the gas was compressed by dynamical pressure of the falling motion to form the HI arch along the field lines.\\
(4) The most strongly compressed gas in the accumulating region at the root of the field lines, a phase transition from HI to H$_2$ occurred to form the Aquila-Serpens molecular complex. \\
(5) The molecular complex and HI arch are separated by a thin layer of phase transition, about which the two phases of gas avoid each other (figure \ref{HIvsCO}).\\
(6) The Parker instability took place in a sinusoidal wave, so that the counter arch to the Aquila Rift is observed at negative latitudes around $(l,b)\sim (300\deg,-30\deg)$ (figures 3 and 13).\\
(7) The physical parameters of the present Aquila Rift are observed as listed in table \ref{tabpara}.

\subsection{Relation to the North Polar Spur and Background Radio Emissions} 

The Aquila Rift is composed of HI, CO and dust clouds, and hence it is a low-temperature object. Therefore, it may neither contain high-temperature plasma to radiate thermal radio emission, nor radiate synchrotron emission, as usually not the case in HI, CO or dust clouds. Also, it is difficult to discriminate the radio continuum emission, if any, originating in the local space within a few hundred parsecs associated with the Aquila Rift. Hence, we did not use radio continuum data.

The Aquila Rift is apparently located near the root of the North Polar Spur. The NPS radiates strong radio continuum emission (Sofue et al. 1979; Haslam et al. 1982), and its brightest ridge runs almost perpendicular to the Aquila Rift at $l\sim 20-30\deg$ and $b=5-20\deg$. The magnetic field direction inferred from radio synchrotron polarization is parallel to the NPS ridge (WMAP: Bennett et al. 2013; Sun et al. 2014, 2015; Planck Collaboration 2015c). On the other hand the magnetic field direction from polarization of FIR dust emission is parallel to the Aquila Rift (figure 10; Planck Collaboration 2015a,b), which is nearly perpendicular to the NPS.

In our recent paper (Sofue 2015) we derived a firm lower limit of the distance to the NPS to be $1.0\pm 0.2$ kpc by analyzing soft X-ray absorption by local molecular clouds. Sun et al. (2014) showed that the radio emitting NPS is located farther than $2-4$ kpc from Faraday screening analysis. 

From the different distances and perpendicular magnetic fields of the NPS and Aquila Rift, these two objects are separate structures, not related physically, but are apparently superposed on the sky. 
Besides NPS, most of radio continuum features at $l\sim 330\deg - 30\deg$ and $b\sim 0-60\deg$ are also emissions from the galactic halo and/or high-energy shells and bubbles expanding from the Galactic Center (Jones et al. 2012; Croker et al. 2014; Sofue et al. 2016) located at distances of several kpc, far beyond the Aquila Rift.

\subsection{Aquila Rift vs HI shells and loops}

Arch structures of neutral gas similar to the Aquila Rift are observed in various places of the Galaxy in HI channel maps (Kalberla et al. 2003). Some are observed as partial HI shells, loops, filaments and/or worms (Heiles 1979, 1984), or as helical structures (Nakanishi et al. 2016). The HI shells, as the naming suggests, are usually interpreted as tangential projections of expanding front of spherical bubbles. 

However, the Aquila Rift shows a more open structure than the currently known HI shells and loops. It crosses the galactic plane toward the negative latitudes, as shown in the 3D plots and fitted lines in figures \ref{xyz}, exhibiting a sinusoidal behavior about the galactic plane with a wavelength as long as $\lambda \sim 2.5$ kpc. It is also associated with molecular complex at the root near the galactic plane.

\subsection{Similarity to extragalactic dust arches}

\begin{figure} 
\begin{center} 
\includegraphics[width=7cm]{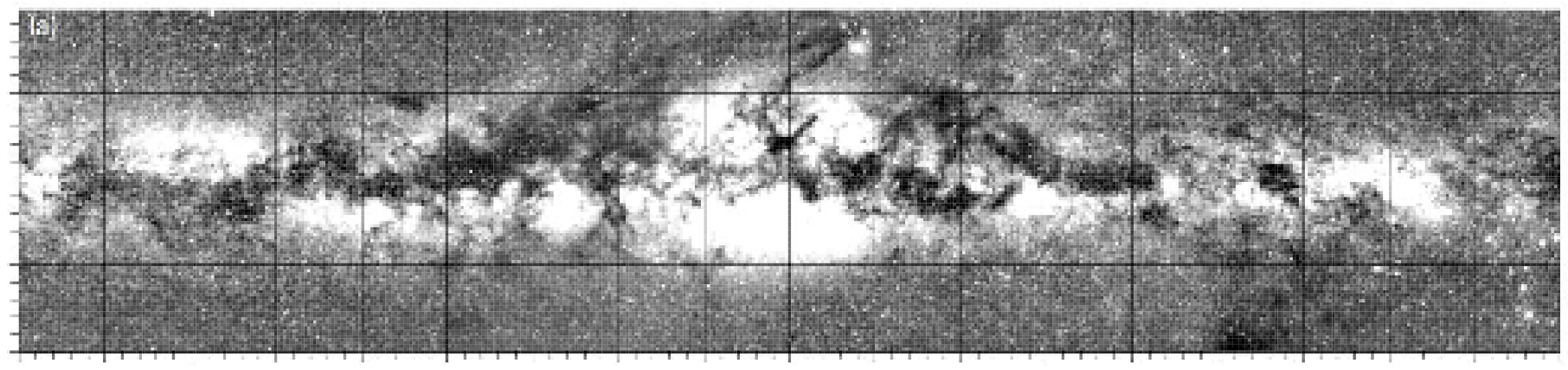}   \\
$90\deg$ ~~~~~~~~~~~~~~~~~~~~~~~~~~~$l=0\deg$~~~~~~~~~~~~~~~~~~~~~~~~~~~~~$270\deg$ \\
\vskip 3mm\includegraphics[width=7cm]{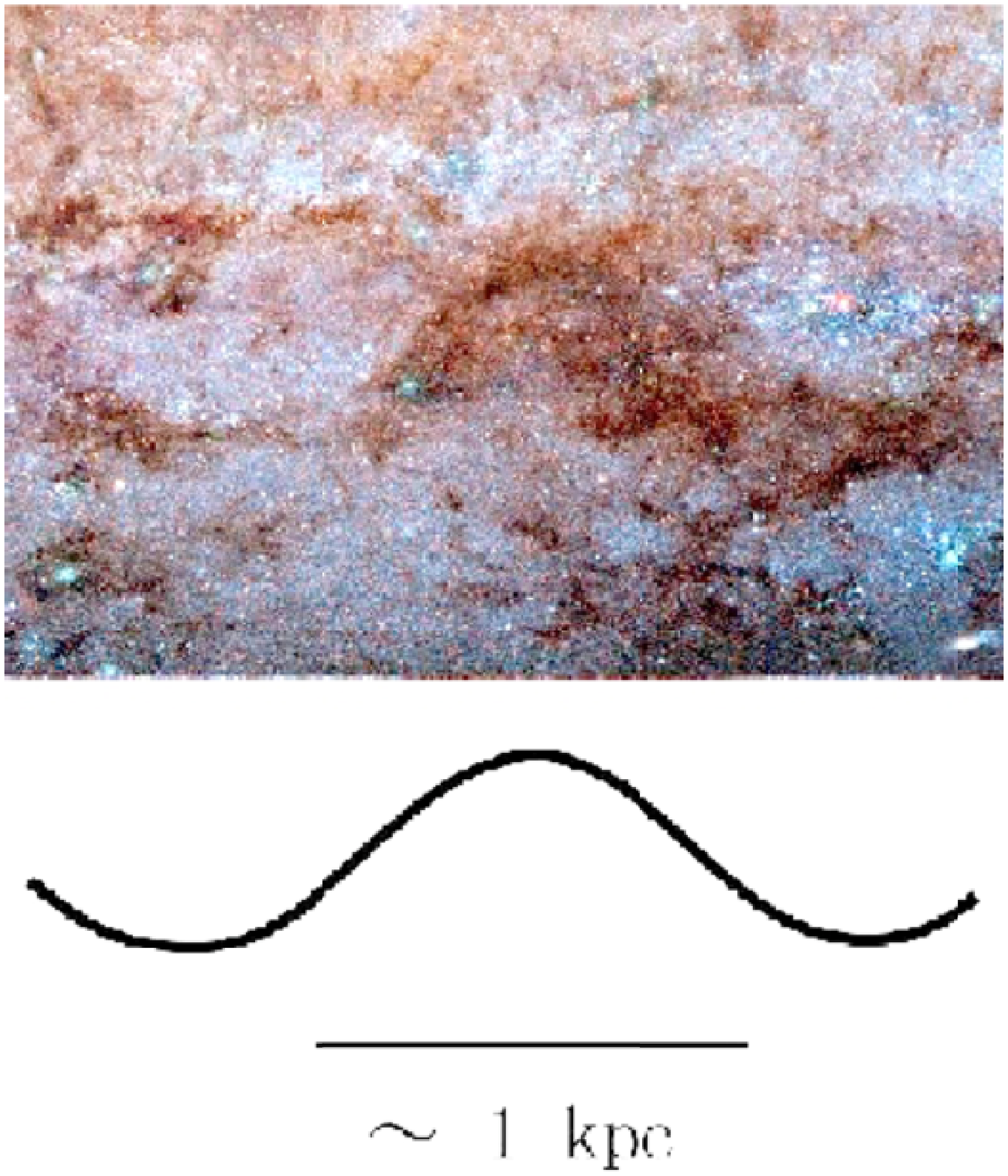}   

\end{center}
 {\caption{The Aquila Rift in dark clouds associated with molecular clouds (Dame et al. 2001) is compared with a dust arch in the spiral galaxy NGC 253, suggesting a sinusoidal Parker instability of wavelength $\sim 2$ kpc (Sofue et al. 1994). (Photo and courtesy: http://apod.nasa.gov/apod/ap110519.html)}
}
\label{N253}
\end{figure}

A survey of extraplanar dust structures in the galactic disk of the spiral galaxy NGC 253 has revealed numerous interstellar dust arches, which are interpreted  as due to inflating neutral gas by the Parker instability (Sofue et al. 1994). Figure \ref{N253} shows an example of such sinusoidal dust arches found in NGC 253.

The Aquila Rift is similar to the extragalactic wavy dust arches in morphology and sizes. Sofue et al. (1994) estimated energetics of dust arches in NGC 253 and showed that the magnetic energy of an arch is on the order of $\sim 10^{50-51}$ erg, comparable to that of the Aquila Rift.  Thus, we may consider that the Aquila Rift is a nearest case of such wavy arches of magnetized neutral gas in galactic disks.

\section{Conclusion}

 {Three dimensional structure of the Aquila Rift of magnetized neutral gas was investigated by analyzing HI and CO line kinematical data. By applying the $dv/dl$ method to the HI velocity data, the HI arch was shown to be located at is $\sim 250$ pc from the Sun. The main ridge of the HI arch emerges at $l\sim 30\deg$ toward positive latitudes, reaching altitudes as high as $\sim 500$ pc above the plane at $l\sim 330\deg$, and returns to the galactic plane at $l\sim 270\deg$. The ridge also extends to negative latitudes from $\sim(30,0)$ to $(90\deg,-50\deg)$.}

 {The extent of arch at positive latitudes is $\sim 1$ kpc and radius is $\sim 100$ pc. The eastern root is associated with the giant molecular cloud complex of Aquila-Serpens, which is the main body of the optically defined Aquila Rift.}

 The masses of the HI and molecular gases in the arch are estimated to be $M_{\rm HI}\sim 1.4\times 10^5\Msun$ and $M_{\rm H_2}\sim 3\times 10^5\Msun$. Gravitational energies to lift the gases to their heights are estimated to be on the order of $E_{\rm grav: HI}\sim 1.4\times 10^{51}$ and $E_{\rm grav: H_2}\sim 0.3\times 10^{51}$ erg, respectively. 

 {Magnetic field is aligned along the HI arch, and the strength is measured to be $B\sim 10\ \muG$ using Faraday rotation measures of extragalactic radio sources. The arch's magnetic energy is estimated to be $E_{\rm mag}\sim 1.2\times 10^{51}$ erg.}

 From the sinusoidal shape of the HI ridge and magnetic flow lines on the sky, we proposed a possible MHD mechanism of formation of the Aquila Rift. It may be produced by Parker instability occurring in the magnetized galactic disk, and the wavelength is estimated to be $\lambda \sim 2.5$ kpc and amplitude $\sim 500$ pc. The magnetic field lines from MHD simulations projected on the sky can well reproduce the HI arch (figure 13).

\section*{Acknowledgments}
The authors are indebted to the authors of Kalberla et al. (2005), Dame et al. (2005), and Taylor et al. (2009) for the archival data. They also thank Dr. K. Ichiki, Nagoya University, for his help during data analysis about rotation measures. 
\\
\noindent{\bf References}
\\
 
\def\r{\hangindent=1pc  \noindent}   
 
\r  Bennett, C.~L., Larson, D., Weiland, J.~L., et al.\ 2013, ApJS, 208, 20 

\r  Clark S.~E., Peek J.~E.~G., Putman M.~E., 2014, ApJ, 789, 82 

\r Cox, A. N., ed. 2000, in 'Allen's Astrophysical Qantities' 4th edition, Springer, Heidelberg, Ch. 23.

\r   Crocker R.~M., Bicknell G.~V., Taylor A.~M., Carretti E., 2015, ApJ, 808, 107  

\r Dame, T. M., Hartman, D., Thaddeus, P. 2001, ApJ 547, 792.  

\r   Dobashi, K., Uehara, H., Kandori, R., et al.\ 2005, PASJ, 57, 1

\r  Dalgarno A., McCray R.~A., 1972, ARA\&A, 10, 375 
 
\r   Dzib, S., Loinard, L., Mioduszewski, A.~J., et al.\ 2010, ApJ 718, 610  

\r   Egger R.~J., Aschenbach B., 1995, A\&A, 294, L25 

\r  Elmegreen B.~G., 1982, ApJ, 253, 655 

\r  Foster, T., Kothes, R., \& Brown, J.~C.\ 2013, ApJ.L., 773, L11 
 
\r  Lallement R., Snowden S.~L., KUNTZ K., Koutroumpa D., Grenier I., Casandjian J.-M., 2016, HEAD, 15, 110.14 

\r  Hanawa T., Nakamura F., Nakano T., 1992, PASJ, 44, 509 

\r   Heiles C., 1979, ApJ, 229, 533  

\r   Heiles, C.\ 1984, ApJS 55, 585 

\r  Heiles C., 1998, LNP, 506, 229 

\r Jenkins, E.~B.\ 2013, ApJ 764, 25

\r  Jones, D.~I., Crocker, R.~M., Reich, W., Ott, J., \& Aharonian, F.~A.\ 2012, ApJL 747, L12  

\r   Lee S.~M., Hong S.~S., 2011, ApJ, 734, 101

\r  Kalberla, P.~M.~W., Burton, W.~B., Hartmann, D., et al.\ 2005, AA 440, 775

\r  Kawamura, A., Onishi, T., Mizuno, A., Ogawa, H., \& Fukui, Y.\ 1999, PASJ, 51, 851 

\r Lee S.~M., Hong S.~S., 2011, ApJ, 734, 101

\r Machida M., et al., 2009, PASJ, 61, 411

\r Machida M., Nakamura K.~E., Kudoh T., Akahori T., Sofue Y., Matsumoto R., 2013, ApJ, 764, 81 

\r Mathewson, D. S. and Ford, V. L. 1970 MNRAS 73, 139. 

\r   Matsumoto R., Hanawa T., Shibata K., Horiuchi T., 1990, ApJ, 356, 259

\r  Mouschovias T.~C., Kunz M.~W., Christie D.~A., 2009, MNRAS, 397, 14  

\r  Nozawa S., 2005, PASJ, 57, 995 

\r  Parker E.~N., 1966, ApJ, 145, 811 
  
\r   Planck Collaboration, et al., 2015a, AA 576, A104 

\r   Planck Collaboration, et al., 2015c, A\&A, 576, A105 
 
\r  Puspitarini, L., Lallement, R., Vergely, J.-L., \& Snowden, S.~L.\ 2014, AA 566, A13 
 
\r   Reis W., Corradi W.~J.~B., 2008, A\&A, 486, 471 

\r   Rodrigues L.~F.~S., Sarson G.~R., Shukurov A., Bushby P.~J., Fletcher A., 2016, ApJ, 816, 2 

\r   Santill{\'a}n A., Kim J., Franco J., Martos M., Hong S.~S., Ryu D., 2000, ApJ, 545, 353  

\r  Santos F.~P., Corradi W., Reis W., 2011, ApJ, 728, 104 

\r Sofue, Y. 2006, PASJ 58, 335.

\r  Sofue Y., 2015, MNRAS, 447, 3824

\r Sofue Y., Habe A., Kataoka J., Totani T., Inoue Y., Nakashima S., Matsui H., Akita M., 2016, MNRAS, 459, 108   

\r Sofue, Y. and Reich, W. 1979 AAS 38, 251  
 
\r Sofue, Y., Wakamatsu, K., and Malin, D. F. 1994, AJ 108, 2102
 
\r [Sun et al.(2014)]{2014MNRAS.437.2936S} Sun, X.~H., Gaensler, B.~M., Carretti, E., et al.\ 2014, \mnras, 437, 2936  

\r  Sun, X.~H., Landecker, T.~L., Gaensler, B.~M., et al.\ 2015, ApJ 811, 40 

\r Sun, X.~H., Gaensler, B.~M., Carretti, E., et al.\ 2014, MNRAS 437, 2936   

\r   Weaver, H.~F.\ 1949, \apj, 110, 190

\r   Taylor A.~R., Stil J.~M., Sunstrum C., 2009, ApJ, 702, 1230

\r   Vidal M., Dickinson C., Davies R.~D., Leahy J.~P., 2015, MNRAS, 452, 656

\end{document}